# Detection of Sleep Apnea-Hypopnea Events Using Millimeter-wave Radar and Pulse Oximeter

Wei Wang[1], Chenyang Li[1], Zhaoxi Chen, Wenyu Zhang, Zetao Wang, Xi Guo, *Member, IEEE*,
Jian Guan[*] and Gang Li[*], *Senior Member, IEEE*

*Abstract*—Obstructive Sleep Apnea-Hypopnea Syndrome (OSAHS) is a sleep-related breathing disorder associated with significant morbidity and mortality worldwide. The gold standard for OSAHS diagnosis, polysomnography (PSG), faces challenges in popularization due to its high cost and complexity. Recently, radar has shown potential in detecting sleep apnea-hypopnea events (SAE) with the advantages of low cost and non-contact monitoring. However, existing studies, especially those using deep learning, employ segment-based classification approach for SAE detection, making the task of event quantity estimation difficult. Additionally, radar-based SAE detection is susceptible to interference from body movements and the environment. Oxygen saturation ($SpO_2$) can offer valuable information about OSAHS, but it also has certain limitations and cannot be used alone for diagnosis. In this study, we propose a method using millimeter-wave radar and pulse oximeter to detect SAE, called ROSA. It fuses information from both sensors, and directly predicts the temporal localization of SAE. Experimental results demonstrate a high degree of consistency (ICC=0.9864) between AHI from ROSA and PSG. This study presents an effective method with low-load device for the diagnosis of OSAHS.

## I. Introduction

Obstructive sleep apnea-hypopnea syndrome (OSAHS) is one of the most common sleep-related breathing disorders in the world [1, 2]. It is characterized by recurrent collapse of the upper airway and reductions in ventilation, leading to a decrease in the sleep quality [3]. Sleep apnea-hypopnea events (SAE) typically include central, obstructive, mixed apnea (CA, OA, MA) and hypopnea (H). Severe OSAHS may lead to various associated comorbidities, such as hypertension, diabetes, anxiety, and stroke, significantly impacting human health and quality of life. According to review studies, the prevalence of OSAHS among adults worldwide ranges from 9% to 38%, and it may exceed 78% in the elderly population [4].

[1] Co-first authors
* Corresponding authors: Jian Guan (guanjian0606@sina.com), Gang Li (gangli@tsinghua.edu.cn)

This work was supported by National Natural Science Foundation of China under Grants 61925106.

Wei Wang, Xi Guo and Gang Li are with the Department of Electronic Engineering, Tsinghua University, Beijing 100084, China.

Chenyang Li and Jian Guan are with the Department of Otolaryngology-Head and Neck Surgery & Center of Sleep Medicine, Shanghai JiaoTong University school of medicine Affiliated Sixth People's Hospital, Shanghai Key Laboratory of Sleep Disordered Breathing, Shanghai 200233, China.

Zhaoxi Chen, Wenyu Zhang and Zetao Wang are with the Beijing Qinglei Technology Co. Ltd., Beijing 100089, China.

Polysomnography (PSG) serves as the gold standard for diagnosing OSAHS. It involves the analysis of sleep disturbances based on multiple physiological signals recorded during sleep, including airflow, respiratory efforts, oxygen saturation ($SpO_2$), electrocardiogram (ECG), etc [5]. Typically, PSG requires patients to visit the hospital and wear various contact-based sensors during the sleep. Sleep technologists subsequently annotate SAE based on the physiological signals. The numerous electrodes attached to the body surface can cause discomfort to patients and reduce their sleep quality. Additionally, the 'First-Night Effect' [6] may lead to deviations in a patient's sleep patterns compared to their regular sleep, affecting the diagnostic outcomes.

Radar has been widely used in non-contact respiration monitoring, including measurements of respiration rate (RR) [7, 8] and tidal volume [9, 10]. In recent years, many studies have employed radar technology for SAE detection using empirical feature-based methods [11, 12] and deep learning methods [13, 14]. The selection of empirical features relies on domain expertise and experience, thereby potentially resulting in subjectivity and limitations in feature selection. Existing deep learning methods typically classify divided segments as either normal or pertaining to specific event categories [13, 14]. The fixed temporal length of segments introduces a challenge in accurately estimating the number of events, as it may result in the inclusion of multiple short events within a single segment or the segmentation of long events into multiple segments. Segment division strategies also demand careful consideration [15]. Faster R-CNN is widely used in object detection, which directly outputs bounding boxes and their categories [16]. The variant of it has also been proposed for temporal localization of specific actions in videos [17], offering valuable insights for radar-based SAE detection. Although radar-based SAE detection achieves satisfactory performance in ideal lab environments, it may still experience misjudgments when influenced by body movements and environmental interference.

Recent studies have demonstrated that oxygen saturation can offer valuable information for detecting SAE, but it also has some limitations [18]. First, pulse oximetry is susceptible to poor peripheral arterial blood flow, so vasoconstriction and hypotension may cause $SpO_2$ artifacts. Variations in the structure and quantity of hemoglobin can also cause artificial increases or decreases that are not due to SAE. Additionally, some forms of sleep-disordered breathing may not result in oxygen desaturation, such as central sleep apnea [18].

Compared to PSG, radar enables non-contact monitoring and pulse oximeter also has high wearing comfort. Both sensors are cost-effective and portable. Inspired by these developments, we propose a method using millimeter-wave radar and pulse oximeter for the detection of SAE, called ROSA. It includes a radar-based SAE detection network using R-CNN structure, called RASA R-CNN, and a fusion algorithm to integrate SpO$_2$ information. RASA R-CNN takes the pre-processed spectrograms of radar signals as input and outputs the temporal localization of SAE during sleep. We also estimate the AHI of subject for preliminary diagnosis of OSAHS. Experimental results demonstrate that ROSA exhibited excellent performance in the detection of SAE.

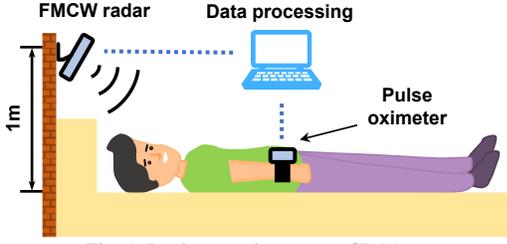

Fig. 1. Implementation scene of ROSA

## II. METHODOLOGY

Our method has two inputs: radar and SpO$_2$ signals. The radar is placed directly above the head of the bed (Fig. 1). Its elevation angle is adjusted to ensure that the line of sight (LOS) faces the chest of the subject. Subjects are additionally instructed to wear a fingertip pulse oximeter to record SpO$_2$ signals.

### A. Radar Signal Pre-processing

In pre-processing, the received signals of FMCW radar are transformed into three spectrograms with different physical significance. These spectrograms portray the spatial and temporal distributions of the respective physical quantities.

First, the received signals are mixed with the transmitted signals to obtain the beat signals $x[\tau, t]$ [10], where $\tau$ denotes the fast time (timestamps within a chirp) and $t$ denotes the slow time (timestamps of different chirps). We perform the windowed fast Fourier transform on fast time of $x[\tau, t]$ to achieve the range-slow-time matrix $R[r, t]$, where $r$ denotes the range.

Then we conduct a series of filtering operations along the slow-time dimension of $R[r, t]$. The high-pass filtering with a cut-off frequency of $f_c$ = 5Hz is employed to extract signals which reflect human body movements. The band-pass filtering with a passband ranging from 0.1Hz to 5Hz is employed to extract respiration-related signals. The first spectrogram $x_M[r, t]$, denoting the power associated with the body movement, can be obtained according to the high-frequency signals. The second spectrogram $x_B[r, t]$, denoting the power of breathing, can be obtained according to the low-frequency signals. The chest movements resulting from respiration can induce the Doppler frequency in the received signals. We perform the Doppler analysis on the low-frequency signals to extract the two-dimensional distribution of the doppler frequency principal component over range and time, denoted as $x_D[r, t]$. Fig. 2 shows an example of three pre-processed spectrograms aligned with airflow, respiratory effort and SpO$_2$. Corresponding changes are evident in our spectrograms concurrent with the occurrence of OA. The three pre-processed spectrograms are concatenated along the channel dimension to form a new spectrogram with three channels, which is then inputted into the proposed RASA-RCNN.

### B. RASA R-CNN Structure

The proposed RASA R-CNN follows the detection paradigm of Faster R-CNN [16], as shown in Fig. 3. Object detection is committed to detect 2D spatial regions containing specific objects. In this study, the goal of SAE detection is to detect 1D temporal segments containing specific events, which can be regarded as 1D object detection. Each detected segment is characterized by four parameters: $y$, $p$, $t_{start}$ and $t_{end}$, denoting the category, score, start time and end time, respectively.

RASA R-CNN takes the concatenated spectrogram as input and outputs the parameters of all detected segments. It contains two stages similar to Faster R-CNN. First, we use a ResNet backbone with feature pyramid structure. The 2D feature map is compressed along the range dimension and subsequently inputted into a Segment Proposal Network (SPN), which is a 1D variant of the Region Proposal Network (RPN) [16]. SPN generates anchor segments of various scales as candidate segments at each time step. It then proposes Segments of Interest (SoIs) for the next stage and regresses their boundaries. Second, the segment proposals are aligned using a 1D variant of the RoIAlign layer in Mask R-CNN [19]. A deep neural network (DNN) is used to predict the category of SoIs and further regress their boundaries. Finally, RASA R-CNN outputs all detected segments represented by parameters introduced above.

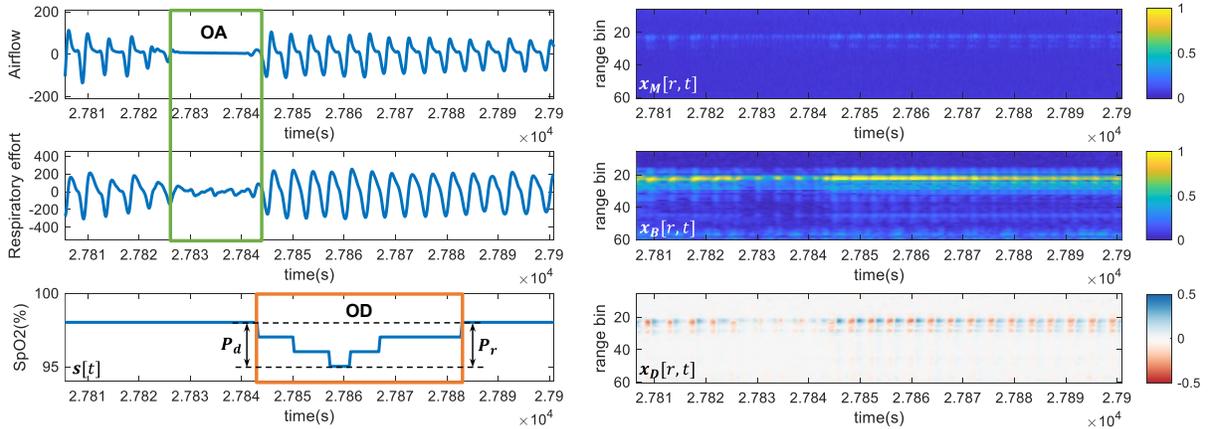

Fig. 2. Pre-processed spectrograms aligned with physiological signals

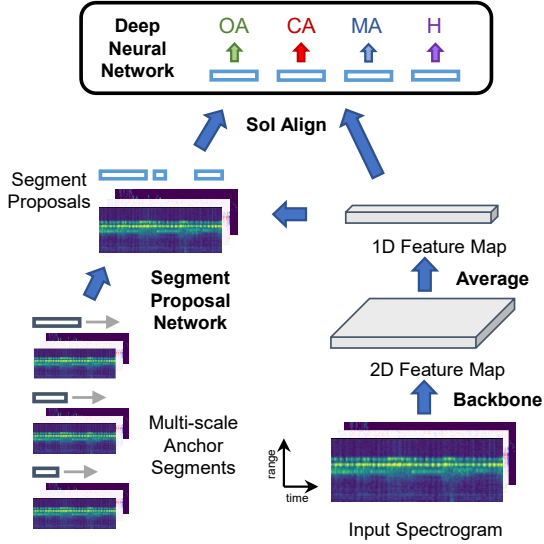

Fig. 3. RASA R-CNN architecture

*C. Fusion of Radar and Oxygen Saturation*

SAE typically causes recurrent oxygen desaturation [20]. The proposed fusion algorithm adjusts the score of the radar-detected segments according to the $SpO_2$ features.

We denote the $SpO_2$ signals as $s[t]$. For each detected segment with a start time denoted as $t_{start}$, feature extraction is performed on the $SpO_2$ signals $s[t_{start}:t_{start}+\Delta t]$, where $\Delta t$ is set to 1 minute. Specifically, we extract the decrease percentage $P_d$ of the first oxygen desaturation (OD) exceeding 3%, along with the subsequent oxygen saturation rise percentage $P_r$ (shown in Fig. 2). If there is no OD greater than 3%, we extract the percentage of the maximum OD.

For detected segments with significant OD, we consider it highly probable to be SAE and increase the score generated by RASA R-CNN. Conversely, for segments with minimal OD, we interpret them as probable false positives and reduce the score of such segments. The detailed fusion algorithm can be expressed as:

$$p' = \begin{cases} \alpha * p + (1-\alpha), & \text{if } P_d \geq T_1 \text{ or } P_r \geq T_1 \\ \beta * p, & \text{if } P_d < T_2 \text{ and } P_r < T_2, \\ p, & \text{else} \end{cases} \quad (1)$$

where $\alpha$ and $\beta$ are weighting coefficients, $T_1$ and $T_2$ are learnable thresholds. The aforementioned fusion algorithm can be seen as a weighted summation of score. The detected segments with significant OD are assigned a score of 1, while those exhibiting minimal OD are assigned 0. Subsequently, we summarize the scores from RASA R-CNN and $SpO_2$ information with the weighting coefficients to obtain the final score, denoted as $p'$. We set $\alpha$=0.5 and $\beta$=0.6 empirically in our study. The optimal thresholds in (1) are determined using grid search on the training and validation set. The objective of grid search is to maximize the Intraclass correlation coefficient [21] (ICC) between the AHI from ROSA and PSG.

Table I. Configurations of FMCW radar

| Parameter | Configuration |
|---|---|
| Start frequency | 60 GHz |
| Sweep bandwidth | 3 GHz |
| Frame rate | 50 Hz |
| Samples per chirp | 256 |

Table II. Subjects demographics and sleep-related variables

| Group | Num (M/F) | Age (years) | BMI (kg/m$^2$) | TST (h) | AHI (events/h) |
|---|---|---|---|---|---|
| Healthy | 27(6/21) | 27.1±8.6 | 21.2±2.5 | 7.1±1.7 | 2.3±1.0 |
| Mild | 32(16/16) | 28.4±7.3 | 22.8±2.4 | 6.7±1.6 | 8.1±2.2 |
| Moderate | 16(14/2) | 37.1±8.9 | 26.0±3.6 | 7.4±1.2 | 21.9±3.9 |
| Severe | 25(21/4) | 44.7±11.1 | 27.2±3.0 | 7.3±1.0 | 57.2±18.1 |

Age, BMI, TST and AHI are presented as mean ± standard deviation; M=Male; F = Female.

### III. EXPERIMENTAL RESULTS

*A. Data Collection*

We used a FMCW radar system based on Infineon-BGT60TR13C and a ChoiceMMed MD300W628 pulse oximeter to simultaneously monitor subjects who undergo PSG during sleep (Fig. 1). Infineon-BGT60TR13C is a 60GHz Antenna-In-Package (AiP) radar equipped with one transmitter and three receivers. We utilized a single receiver for the purpose of our study. The radar parameters are configured as shown in Table I. PSG provides the ground truth of SAE. All PSG data were annotated by sleep technologists, according to the AASM manual for the scoring of sleep and associated events [22].

All data for this study were collected from the Shanghai JiaoTong University school of medicine Affiliated Sixth People's Hospital. This study was conducted in accordance with the Declaration of Helsinki and the study protocol was approved by the Ethics Committee of Shanghai JiaoTong University Affiliated Sixth People's Hospital (2023-030-[1]). The study was registered at the United States Clinical Trial Registry (No. NCT06038006). All subjects provided informed consent. The basic information of the subjects is shown in Table II.

*B. Implementation Details*

The proposed RASA R-CNN is implemented on Pytorch. We train the network by stochastic gradient descend (SGD) optimizer with momentum for 80 epochs. The learning rate is set with reference to the cosine annealing learning rate. The weighted cross entropy loss is used to deal with data imbalance. We apply a 4-fold cross-validation technique splitting the data based on individual subjects, and report the performance on the test set. In the training stage, input data consists of half-hour segments to support enough batch size. During the inference stage, the network processes the whole night segment directly.

*C. Efficacy of Radar and Oxygen Saturation Fusion*

The fusion of radar and $SpO_2$ signal serves to mitigate false positives and enhance the confidence of accurate event detection. The optimal thresholds determined through Grid Search were $T_1$=4% and $T_2$=2%. The comparative analysis of results with $SpO_2$-only, radar-only (RASA R-CNN) and the fusion of radar and $SpO_2$ (ROSA) is detailed in Table III. Particularly, the results with SpO$_2$-only are given by using the 3% oxygen desaturation index (ODI$_3$) in OSAHS diagnosis [20]. The results demonstrate that the overall detection performance exhibits superiority with the fusion of radar and SpO$_2$ signal.

Table III. Detailed results of SAE detection

| Method | AP$_{0.5}$ (%) | ICC | Diagnostic thresholds | Sensitivity (%) | Specificity (%) | Accuracy (%) | Kappa |
|---|---|---|---|---|---|---|---|
| ODI$_3$ | / | 0.9064 | 5 events/h | 73.97 | **100.00** | 81.00 | 0.6055 |
| | | | 15 events/h | 85.37 | 98.31 | 93.00 | 0.8526 |
| | | | 30 events/h | 88.00 | 98.67 | 96.00 | 0.8904 |
| RASA R-CNN | 69.91 | 0.9599 | 5 events/h | **97.26** | 62.96 | 88.00 | 0.6642 |
| | | | 15 events/h | 92.68 | 94.92 | 94.00 | 0.8760 |
| | | | 30 events/h | 88.00 | 97.33 | 95.00 | 0.8649 |
| ROSA | 74.36 | 0.9864 | 5 events/h | 93.15 | 92.59 | **93.00** | 0.8284 |
| | | | 15 events/h | 92.68 | **100.00** | 97.00 | **0.9373** |
| | | | 30 events/h | **96.00** | 97.33 | 97.00 | 0.9211 |

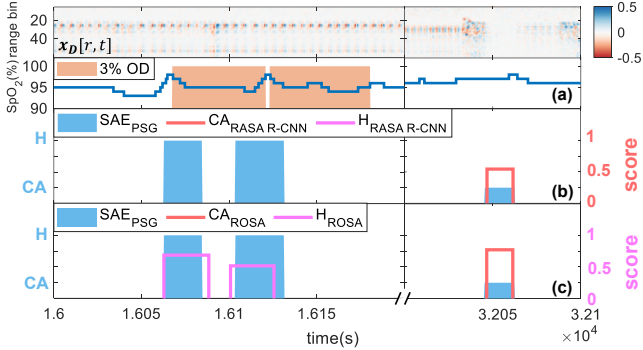

Fig. 4. SAE detection example: (a) SpO$_2$-only; (b) radar-only; (c) fusion. As shown in the figure, the method with SpO$_2$-only misses the central apnea (CA) event, while that with radar-only misses two hypopnea (H) events. With the fusion of radar and SpO$_2$, all SAE can be correctly detected.

The SAE detected by ROSA are compared with those annotated by sleep technologists to evaluate the performance of our method. The Average Precision (AP) is used as the primary metric. In particular, AP$_{0.5}$ indicates the AP when the Intersection over Union (IoU) between predicted and ground truth bounding boxes reaches 0.5. Fig. 4 presents an example of SAE detection during a subject's sleep. The detailed results of SAE detection are listed in Table III. ROSA demonstrates excellent detection performance, achieving the 74.36% on AP$_{0.5}$. This significant result provides support for subsequent estimation of the AHI and the diagnosis of OSAHS.

### D. Comparison of AHI between ROSA and PSG

AHI serves as a crucial metric for diagnosing and quantifying the severity of OSAHS. It delineates the average number of SAE per hour of sleep. The definition of AHI is:

$$\text{AHI} = (N_{\text{apnea}} + N_{\text{hypopnea}})/\text{TST}, \qquad (2)$$

where $N_{\text{apnea}}$ is the number of apnea events, $N_{\text{hypopnea}}$ is the number of hypopnea events, and TST, measured in hours, is the total sleep time provided by PSG.

We can estimate the value of AHI for each subject according to the SAE detected by ROSA. ICC is used as the primary metric to compare the agreement between the AHI estimated from ROSA and that provided by PSG. The comparison of the estimated and true AHI is shown in Fig. 5. It demonstrates that ROSA can have a high agreement (ICC=0.9864) with PSG. We also assess the diagnostic performance of ROSA using AHI thresholds of 5, 15 and 30 events/h. It is evident that ROSA exhibits outstanding diagnostic performance for OSAHS, exceeding 90% in sensitivity, specificity, and accuracy across all thresholds. These results demonstrate the effectiveness of ROSA in OSAHS diagnosis and highlight its clinical value.

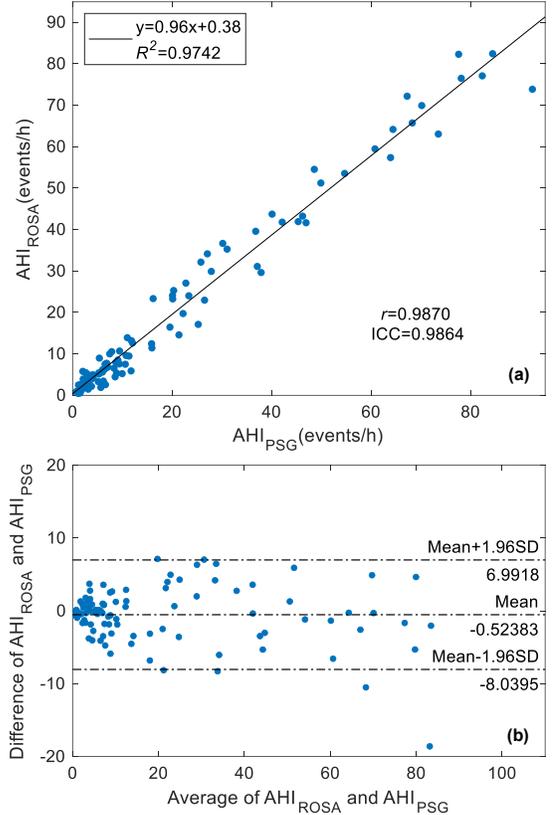

Fig. 5. The comparison between the results obtained from ROSA and PSG: (a) scatter plot depicting the correlation; (b) Bland-Altman plot.

## IV. CONCLUSION

In this study, a method called ROSA is proposed for the detection of SAE using millimeter-wave radar and pulse oximeter. This method can effectively fuse the information from radar and oximeter to improve the performance of SAE detection. Furthermore, it shows the capability to directly provide the temporal localization of SAE without any post-processing. Experimental results demonstrate that ROSA exhibits a high agreement with PSG and showcases commendable diagnostic performance. This study introduces a more convenient and reliable option for the diagnosis of OSAHS.